\documentclass[letter]{aa}  

\usepackage{graphicx}
\usepackage{txfonts}
\usepackage{amsmath,siunitx}
\usepackage[]{hyperref}

\begin{document} 

\renewcommand{\linenumbers}{}
\renewenvironment{linenumbers}{}{}

   \title{Radio emission from a massive node of the cosmic web}

   \subtitle{A discovery powered by machine learning}

   \author{C. Stuardi\inst{1}, A. Botteon\inst{1}, M. Sereno\inst{2,3}, K. Umetsu\inst{4}, R. Gavazzi\inst{5,6}, A. Bonafede\inst{7,1} \and C. Gheller \inst{1}}

   \authorrunning{C. Stuardi et al.}

   \institute{INAF - Istituto di Radio Astronomia, Via P. Gobetti 101, 40129 Bologna, Italy\\
              \email{ccstuardi@gmail.com}
              \and
             INAF - Osservatorio di Astrofisica e Scienza dello Spazio di Bologna, via Piero Gobetti 93/3, 40129 Bologna, Italy
             \and
             INFN, Sezione di Bologna, Viale Berti Pichat 6/2, 40127 Bologna, Italy
             \and
             Academia Sinica Institute of Astronomy and Astrophysics (ASIAA), No. 1, Section 4, Roosevelt Road, 10617, Taipei, Taiwan
             \and
            Laboratoire d'Astrophysique de Marseille, Aix-Marseille Univ., CNRS, CNES, Marseille, France
            \and
            Institut d'Astrophysique de Paris, UMR 7095, CNRS \& Sorbonne Université, 98 bis Boulevard Arago, 75014, Paris, France
             \and
             Dipartimento di Fisica e Astronomia, Universit\'{a} di Bologna, Via P. Gobetti 92/3, 40129 Bologna, Italy
             }

   \date{Received September XX; accepted YY}
   

 
  \abstract
   {The recent detection of radio emission extending beyond the scales typically associated with radio halos challenges our understanding of how energy is transferred to the non-thermal components in the outskirts of galaxy clusters, suggesting the crucial role of mass accretion processes. So far, discoveries have relied on the visual identification of prominent clusters within limited samples. Today, machine learning promises to automatically identify an increasing number of such sources in wide-area radio surveys.}
   {We aim to understand the nature of the diffuse radio emission surrounding the massive galaxy cluster PSZ2\,G083.29-31.03, at z=0.412, already known to host a radio halo. Our investigation was triggered by Radio U-Net, a novel machine learning algorithm for detecting diffuse radio emission, which was previously applied to the LOFAR Two Meter Sky Survey (LoTSS).}
   {We re-processed LoTSS (120-168 MHz) data and analyzed archival XMM-Newton (0.7-1.2 keV) observations. We also analyzed optical and near-infrared data from the DESI Legacy Imaging Surveys and asses the mass distribution with weak-lensing analysis based on archival Subaru Suprime-Cam and CFHT MegaPrime/MegaCam observations.}
   {We report the discovery of large-scale diffuse radio emission around PSZ2\,G083.29-31.03, with a projected largest linear size of 5 Mpc at 144 MHz. The radio emission is aligned with the thermal X-ray emission and the distribution of galaxies, unveiling the presence of two low-mass systems, at similar redshifts on either side of the central cluster. The weak lensing analysis supports this scenario, demonstrating the presence of an extended and complex mass distribution.}
   {We propose to interpret the two faint radio sources as connected to the central cluster, thus illuminating the presence of two substructures merging into a massive node of the cosmic web. However, because of uncertainties in redshift and mass estimates, combined with the low resolution required to detect these sources, classification of the two sources as independent radio halos associated with nearby low-mass clusters or even as a mixture of different types of diffuse radio emission cannot be definitively ruled out.}



   \maketitle
%

\section{Introduction}

The presence of diffuse radio emission filling the volume of massive galaxy clusters up to the characteristic radius $R_{200}$\footnote{$R_{200}$ indicates the radius of a spherical volume whose mean density is 200 times the critical density at the cluster's redshift. A similar definition applies to $R_{500}$, while $M_{500}$ is the mass contained in this volume.}, and extending even beyond, has gained significant interest from the radio astronomy community in recent years, particularly due to advances in low-frequency observations \citep{Govoni19, Shweta20, Rajpurohit21, Cuciti22, Botteon22b, Bruno23}. Theoretical studies, supported by state-of-the-art magneto-hydrodynamical simulations, suggest that this kind of emission can be originated by the turbulence injected in the intra-cluster medium (ICM) by the large-scale structure formation process \citep{Beduzzi23,Nishiwaki24}. Whether this type of radio emission exists in all systems but remains below the noise level for most, or if it is only activated in particularly massive and dynamically disturbed systems, remains to be investigated. In this context, a multiwavelength analysis is necessary to understand the connection between the diffuse radio emission and dynamical state.

In this letter, we report the discovery of a large-scale diffuse radio emission around \object{PSZ2\,G083.29-31.03} (a.k.a. \object{MACS J2228.5+2036} or \object{RXJ2228.6+2037}) at z=0.412±0.002 \citep{Amodeo18}, made using a machine learning algorithm. To our knowledge, this is the first radio diffuse emission discovered thanks to machine learning that went unnoticed by human analysis. In Sec.~\ref{sec:discovery} we explain how the radio emission was discovered with the use of Radio U-Net \citep{Stuardi24}. In Sec.~\ref{sec:analysis} we describe radio, X-ray, optical and near-infrared, and weak lensing analysis. In Sec.~\ref{sec:discussion} we compare multiwavelength observations and discuss possible interpretations, and in Sec.~\ref{sec:conclusions} we provide concluding remarks. Appendix~\ref{app} provides additional information on the radio data analysis.

Using a $\Lambda$ Cold Dark Matter cosmological model, with $\Omega_{\Lambda} = 0.7$, $\Omega_{m} = 0.3$, and $H_0 = 70$~km~s$^{-1}$ ~Mpc$^{-1}$, 1$\arcsec$ corresponds to 5.47 kpc at the galaxy cluster's redshift.


\section{Discovery}
\label{sec:discovery}

   \begin{figure}
   \centering
   \includegraphics[width=\linewidth]{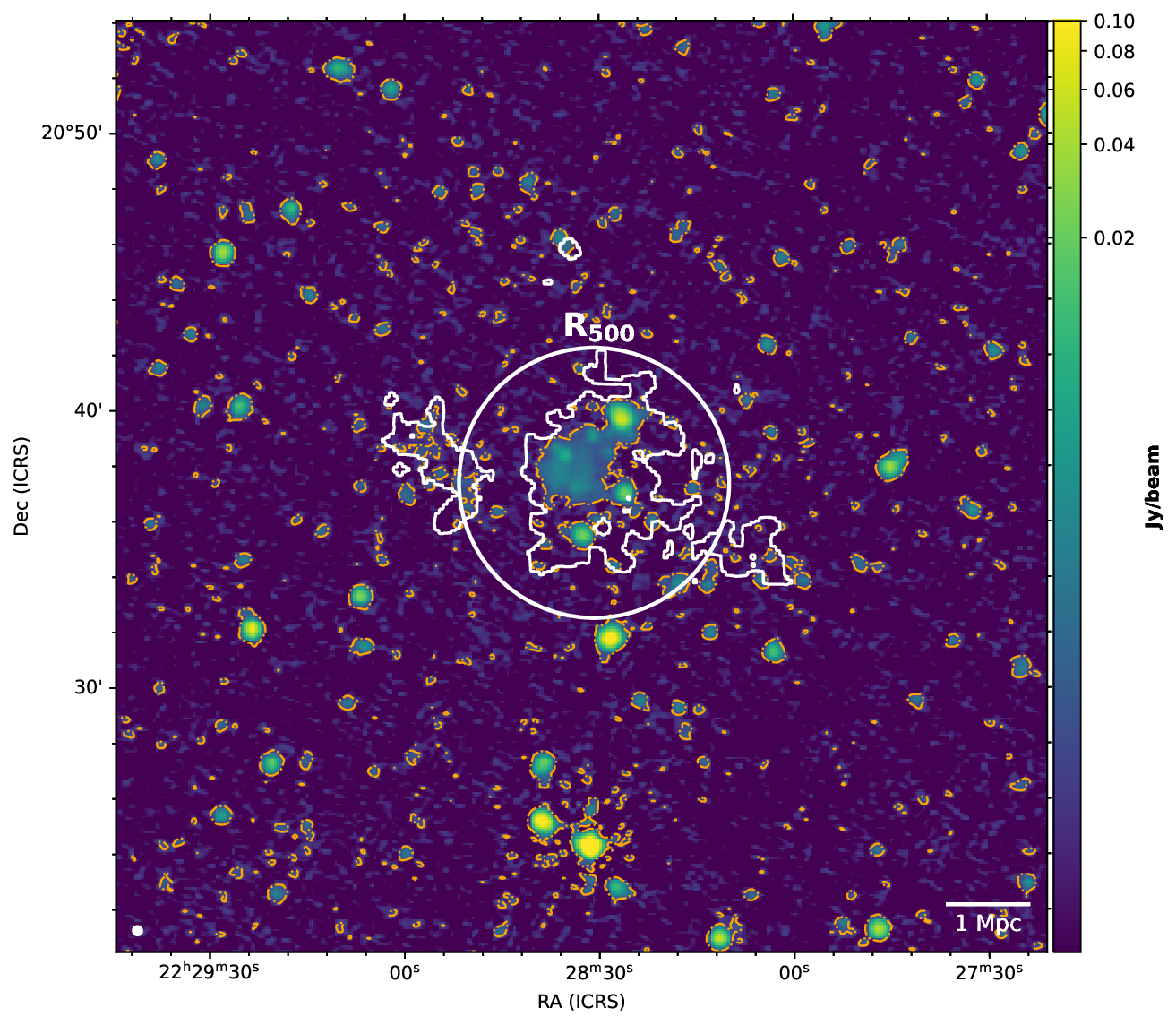}
      \caption{Original LoTSS DR2 image (144 MHz) of PSZ2\,G083.29-31.03, on which Radio U-Net was applied. The restoring beam is shown in white at the bottom left corner and has a full-width half-maximum (FWHM) of 20$\arcsec$. The dash-dotted orange contours show the 3$\sigma$ with $\sigma=0.13$~mJy/beam. The white contours show the segmentation mask created by the network (0.2 in the probability map), while the circle has a radius $R_{500}$ \citep{Jia08}.}
     \label{fig:RadioUnet}
   \end{figure}

      \begin{figure}  
   \centering
   \includegraphics[width=\linewidth]{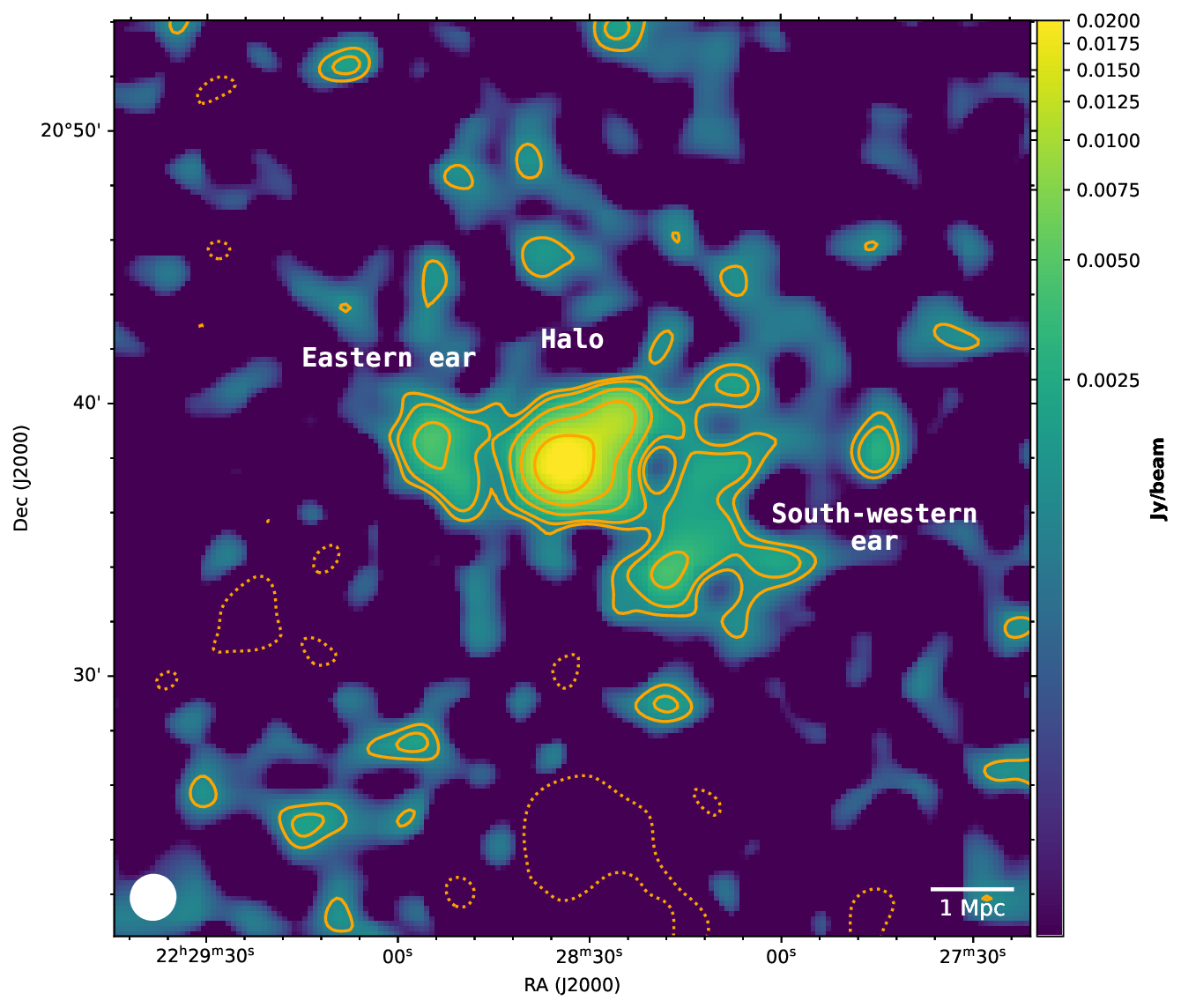}
      \caption{UV-subtracted LoTSS image (144 MHz) with restoring beam of 97$\arcsec\times100\arcsec$. Orange solid contours are [2,3,6,12,24]$\times\sigma$ with $\sigma=0.59$~mJy/beam. The dashed contour marks the -3$\sigma$ level. Strong negative values are reached at the position of two sources which remained poorly subtracted. The shape of the restoring beam is shown in white at the bottom left corner.}
      \label{fig:LOFAR90}
   \end{figure}

With the advent of new-generation wide-area radio surveys, such as the Low Frequency Array \citep[LOFAR,][]{vanHaarlem13} Two Metre Sky Survey \citep[LoTSS,][]{Shimwell17,Shimwell19} and the Evolutionary Map of the Universe \citep[EMU,][]{Norris11}, traditional cataloging methods are becoming less efficient. Machine learning (ML) algorithms are gaining increasing prominence, owing to their ability to automate the detection and classification of sources in large imaging datasets \citep[e.g.][]{Mostert21,Slijepcevic24,Gupta24,Alegre24,Lastufka24}. However, the application of such algorithms to the detection of diffuse radio emission in galaxy clusters and the cosmic web remains poorly explored \citep{Gheller18}.

Recently, we developed a ML tool based on the U-Net architecture, named Radio U-Net, which facilitates the rapid detection and segmentation of diffuse radio sources in large image datasets directly retrieved from archives \citep{Stuardi24}. We tested this algorithm on a cataloged sample of galaxy clusters observed within the LoTSS second data release (DR2) \citep{Botteon22}. The output of the network consists of an image with the same dimensions as the original radio image, where each pixel is assigned a value ranging from 0 to 1, with higher values corresponding to regions of diffuse radio emission. We refer to these outputs as \textit{probability maps}, though their precise definition is provided in \citet{Stuardi24}. Notably, despite Radio U-Net being applied to archival images at 20$\arcsec$ resolution, the network detected emission visible only after smoothing the images to lower resolutions. Additionally, in some cases, the regions highlighted by the probability maps extended beyond those detected at the reference $3\sigma$ level, demonstrating the network's capability to detect diffuse radio emission even below the nominal noise.

One particularly noteworthy cluster included in the sample was PSZ2\,G083.29-31.03. Its probability map revealed the presence of diffuse radio emission extending beyond $R_{500}$, a feature that had been overlooked during the visual inspection of the original image (see Fig.~\ref{fig:RadioUnet}). Radio U-Net clearly identified a peculiar morphology below the noise threshold, triggering a deeper investigation into this system.

PSZ2\,G083.29-31.03 was first listed as a galaxy cluster in the ROSAT All-Sky Survey (RASS) bright source catalog \citep{Voges99} and identified in the optical by \citet{Bade98}. It was then classified as a hot, and massive galaxy cluster by combining Sunyaev-Zel’dovich (SZ) and X-ray observations \citep{Pointecouteau02}. \textit{XMM-Newton} observations estimated the cluster mass $M_{\text{X}, 500} = (1.19\pm0.35)\times10^{15} M_{\odot}$ at $R_{\text{X}, 500} = 1.61\pm0.16$~Mpc, and temperature $T_{500}=8.92^{+1.78}_{-1.32}$~keV \citep{Jia08}. The cluster was listed in the second catalog of Planck Sunyaev Zel’dovich \citep[PSZ2][]{Planck16b} where its mass estimate is $M_{\text{SZ}, 500} = (8.3\pm 0.4)\times10^{14}M_{\odot}$.

A radio diffuse source centered on the cluster X-ray emission was first detected by \citet{Giovannini20} at 1.5 GHz, and classified as a radio halo. They observed radio emission with irregular morphology, with a maximum linear extension of 1.09 Mpc. The presence of the halo was recently confirmed by LOFAR observations \citep{Botteon22}. The morphology of the radio halo at 144 MHz is similar to that observed at 1.5 GHz, although more extended, reaching a linear size of 1.5 Mpc along the northwest-southeast direction.

\section{Data analysis}
\label{sec:analysis}

In the following sections, we present the multi-frequency analysis we have carried out to investigate the origin of the newly detected radio emission around PSZ2\,G083.29-31.03. 

\subsection{Radio data analysis}
\label{sec:radio}

We re-analyzed the LoTSS DR2 observation of the PSZ2\,G083.29-31.03 galaxy cluster already presented by \citet{Botteon22}, where the calibration towards the cluster was optimized with the "extraction+selfcal" method \citep[see][for more details]{vanWeeren21}. This cluster was observed by LOFAR High Band Antennas (HBA) in two slots of 4h which were combined in the P337+21 pointing. The original image at 20$\arcsec$ resolution released by the LoTSS DR2 \citep{Shimwell22} is shown in Fig.~\ref{fig:RadioUnet}.

We removed all sources unrelated to the diffuse emission by making an image with an inner uv-cut at 2865$\lambda$ (corresponding to 0.02$^{\circ}$ and $\sim$400 kpc) and then subtracting the model of this image from the visibilities. Appendix~\ref{app} contains further information on the subtraction procedure and the residual contribution of compact faint sources to the diffuse emission. 

For imaging and prediction procedures, we used the WSClean software \citep[version 3.4,][]{Offringa14,Offringa17}. We produced multi-frequency synthesis (MFS) images with 6 channels and using the multi-scale options with auto-masking (at 3$\sigma$) and auto-threshold (at 2$\sigma$). By imposing a Gaussian taper, we made images of the residual radio emission at different resolutions, shown in Fig.~\ref{fig:LOFAR90} and \ref{fig:X-optical}. We note that the diffuse radio emission is completely deconvolved with this procedure. To align our images with the flux scale of the LoTSS DR2, we have used the same scaling factor computed by \citet{Botteon22}, that is, 1.133. Additional information on all radio images presented in this work is listed in Table~\ref{tab:radioimages}. 

\begin{table}
\tiny
\caption{Details of the LOFAR images (144 MHz).}
\begin{tabular}{cccccc}
  UV-Range & UV-Taper & $\Theta_{FWHM}$ & $\sigma_{rms}$ & UV-Sub. & Fig.  \\
   $\lambda$ & $\arcsec$ & $\arcsec\times\arcsec$ & mJy/beam & & \\
\hline
  >80 & ... & 20$\times$20 & 0.13 & No & \ref{fig:RadioUnet} \\
  >80 & 90 & 97$\times$100 & 0.59 & Yes & \ref{fig:LOFAR90},\ref{fig:X-optical} \\
  >80 & 60 & 69$\times$81 & 0.58 & Yes & \ref{fig:X-optical},\ref{fig:optical-hres-E},\ref{fig:optical-hres-SW} \\
  >2865 & ... & 4.5$\times$10.6 & 0.12 & No & \ref{fig:optical-hres-E},\ref{fig:optical-hres-SW} \\
  >80 & 18 & 27$\times$29 & 0.31 & No & \ref{fig:optical-hres-E},\ref{fig:optical-hres-SW} \\
\hline\end{tabular}
\tablefoot{All images have a Briggs Robust parameter of -0.5. Column 1: image uv-range. Column 2: angular size of the Gaussian uv-taper used. Column 3: restoring beam size. Column 4: root-mean-square (rms) noise. Column 5: flag indicating if point-like sources were subtracted from the image. Column 5: figure reference.}
\label{tab:radioimages}
\end{table} 

There is a clear similarity between the low-resolution image shown in Fig.~\ref{fig:LOFAR90} and the segmentation of the original image produced by Radio U-Net (Fig.~\ref{fig:RadioUnet}). Beyond the emission centered at the cluster position, there are two fainter ‘‘ears’’ extending to the east and the southwest of the cluster. The eastern extension has a roundish morphology and is connected to the brightest part of the central radio halo. The south-western one is instead elongated in the northwest-southeast direction. The morphology of the two ‘‘ears’’ at 69$\arcsec\times81\arcsec$ resolution is shown by the orange contours in Fig.~\ref{fig:X-optical}.

We computed the average surface brightness and the flux density of the two ‘‘ears’’ from the 69$\arcsec\times81\arcsec$ image to avoid contamination from the radio halo. They were computed in two polygonal regions encompassing the $2\sigma$ contours. The east and south-western ‘‘ears’’ have an average surface brightness of 2.01$\pm$0.58 mJy/beam (i.e., 0.32 $\mu$Jy/arcsec$^2$) and 1.79$\pm$0.58 mJy/beam (i.e., 0.28 $\mu$Jy/arcsec$^2$), where the uncertainty is the rms noise of the image. The central radio halo has an average surface brightness that is a factor $\sim$2.7 higher (0.84 $\mu$Jy/arcsec$^2$). We computed a flux density of 20$\pm$3~mJy and 18$\pm$3~mJy for the east and south-western ‘‘ear’’, respectively. The uncertainty on the flux density is computed as in \citet{Botteon22} (more details are given in Appendix~\ref{app}). 

\subsection{X-ray data analysis}

Two available observations of PSZ2\,G083.29-31.03 are present in the \textit{XMM-Newton} archive (ObsIDs: 0147890101, 0827360901). We retrieved and reduced both datasets using the \textit{XMM-Newton} Science Analysis System (SAS) v16.1.0 and the Extended Source Analysis Software (ESAS) data reduction scheme \citep{Snowden08}. We performed a standard data reduction, which started by running \texttt{emchain} and \texttt{epchain} to generate calibrated event files for each observation. Thus, we obtained clean event files by filtering out time periods affected by soft proton flares by using the mos-filter and pn-filter tasks. Photon count images were extracted from the three detects of the EPIC camera in the 0.7--1.2 keV band, and then combined to produce a mosaic of the two ObsIDs.

The X-ray image of PSZ2\,G083.29-31.03 obtained with \textit{XMM-Newton} is shown in Fig.~\ref{fig:X-optical}. A zoom-in with radio contours overlaid is shown in the bottom panel. 

The brightest part of the cluster is elongated along the northwest-southeast direction. However, we notice two clumps of X-ray emission on the east and southwest of the cluster, coincident with the position of the two radio ‘‘ears’’. The eastern clump has a brighter spot at the center and is connected to the central cluster with a faint emission. The distance between the bright spot and the X-ray peak of the cluster is 310$\arcsec$ ($\sim$1.7 Mpc). The fainter clump to the southwest is at a similar projected distance from the central peak.

\subsection{Optical and near-infrared analysis}
\label{sec:optical}

\begin{table}
\tiny
\caption{Neighbor candidate clusters and groups.}
\begin{tabular}{ccccccc}
  Name & R.A. & Dec & $z$ & $M_{500}$ & $R_{500}$ & N$_{gal}$ \\
    &    deg & deg &  & 10$^{14} M_{\odot}$ & Mpc &  \\
\hline
  J222809.9+203528 & 337.04 & 20.59 & 0.416  & 1.7 & 0.75 & 33\\
  J222833.7+203716 & 337.14 & 20.62 & 0.413  & 5.2 & 1.07 & 108\\
  J222854.6+203908 & 337.23 & 20.65 & 0.409  & 1.9 & 0.78 & 35\\
  J222930.8+204011 & 337.38 & 20.67 & 0.416 & 0.6 & 0.43 & 7\\
\hline\end{tabular}
\tablefoot{Entries are extracted from the \citet{WH2024} catalog based on the DESI Legacy Imaging Surveys. Column 1: name. Column 2: Right ascension (J2000). Column 3: Declination (J2000). Column 4: redshift, with an associated uncertainty of 0.01(1+$z$). Column 5: $M_{500}$, with an associated uncertainty of 0.2 dex. Column 6: $R_{500}$. Column 7: number of photometric redshifts used.}
\label{tab:clusters}
\end{table}

The Dark Energy Spectroscopic Instrument (DESI) Legacy Imaging Surveys \citep{DESI_dey+al19} cover more than 20000 square degrees of the extragalactic sky in the optical and near-infrared. \citet{WH2024} identified 1.58 million candidate clusters of galaxies up to $z\sim 1.5$ from data release 9 and 10. Within a matching distance of $5~\text{Mpc}$ at the cluster redshift from the X-ray peak and $\Delta z = \pm 0.02 (1+z)$, we find 4 candidate clusters and groups, arranged as the radio or X-ray distribution, see Table~\ref{tab:clusters} and Fig.~\ref{fig:X-optical}. Based on their precise redshift estimates, the four systems lie within a region of radius 15–20 Mpc. However, the large uncertainties in the photometric redshift and the mass of the systems, derived from their richness by \citet{WH2024}, prevent us from determining whether the systems are physically connected or merely aligned in projection. 

The dynamical mass of $M_{\sigma, 500} = (9.4\pm3.6)\times10^{14}M_{\odot}$ based on the velocity dispersion of the member galaxies \citep{Sereno25} is larger than the richness-based mass of the main clump, but it agrees with the total mass of the four structures.

\subsection{Weak lensing analysis}
\label{sec:wl}

The cluster was targeted for weak lensing (WL) analyses (for a review, see \citealt{Umetsu2020}) by \citet{foe+al12} and \citet{wtg_III_14}, who identified a massive structure with $M_{\text{WL}, 500} = (6.1 \pm 1.3) \times 10^{14}M_{\odot}$ or $M_{\text{WL}, 500} = (9.7 \pm 2.6) \times 10^{14} M_{\odot}$, respectively, as reported by the Literature Catalogue of WL Clusters of galaxies \citep{ser15_comalit_III}.

We conducted a WL analysis of the cluster using the AMALGAM dataset based on archival Subaru Suprime-Cam and CFHT MegaPrime/MegaCam observations (Gavazzi et~al. 2025; Umetsu et~al. 2025, in preparation), produced within the Cluster HEritage project with \textit{XMM-Newton}: Mass Assembly and Thermodynamics at the Endpoint of structure formation collaboration \citep[CHEX-MATE,][]{CHEX-MATE21}. Deep images in the $B, g, V, r, R_C, I_C, z$ bands were used and the shape of galaxies was done on the $R_C$ band using the model fitting capabilities of {\tt SExtractor} \citep{Mandelbaum14}. The attained density of faint background sources is about 24 arcmin$^{-2}$. 

\begin{figure*}     
   \hspace{-2cm}\includegraphics[width=1.2\linewidth]{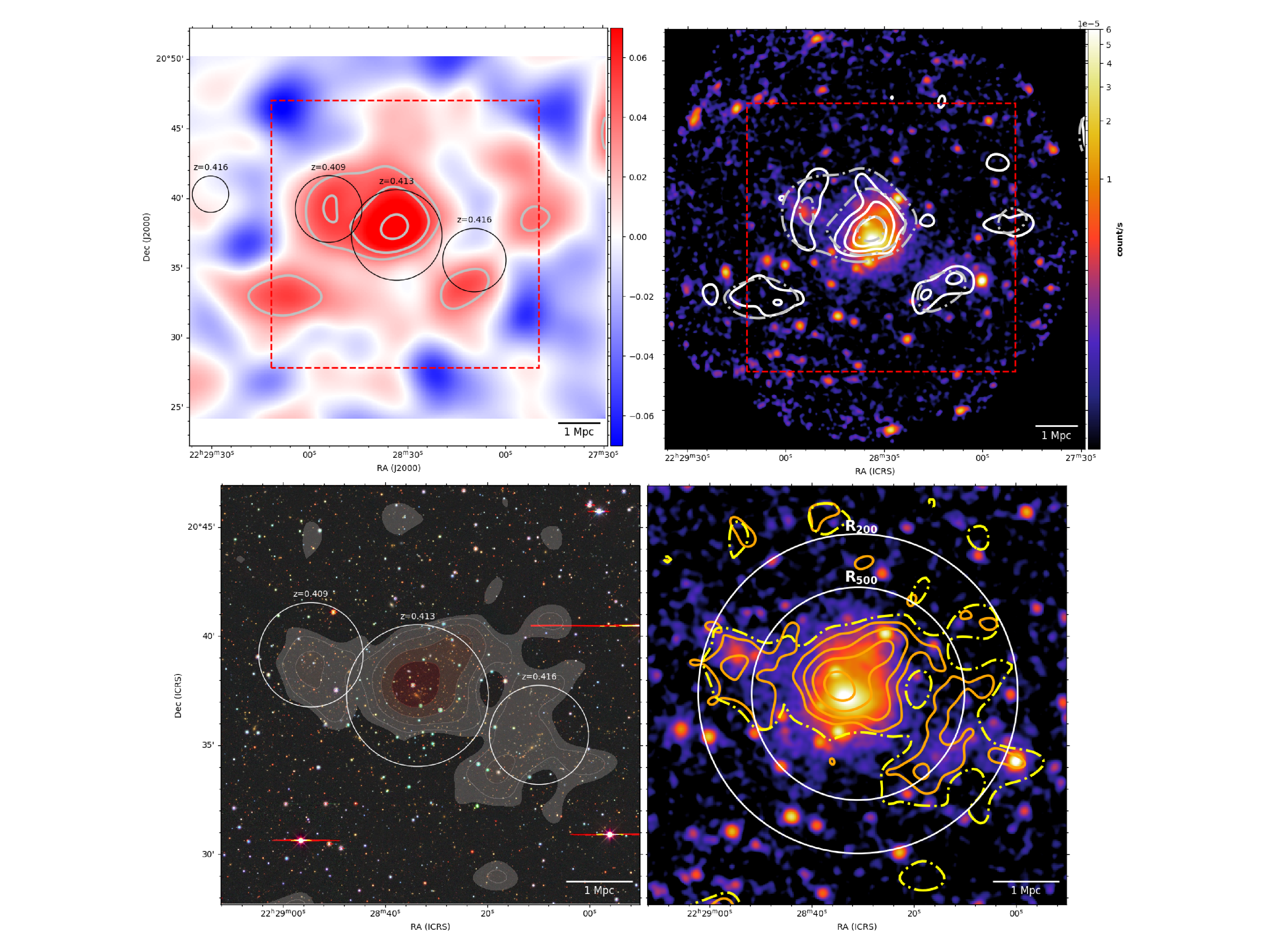}
   \caption{Multi-wavelength view of the PSZ2\,G083.29-31.03 galaxy cluster. Top left panel: WL mass map created with a $3\farcm3$ FWHM Gaussian kernel. Silver contours show the 2$\sigma$, 3$\sigma$, and 5$\sigma$ levels. Black circles mark the systems listed in Tab.~\ref{tab:clusters} and have a radius equal to their estimated $R_{500}$. The dashed red square marks the region of the image shown in the bottom panels. Top right panel: \textit{XMM-Newton} image (0.7-1.2 keV) with a smoothing kernel of 3 pixels. The white contours are 2$\sigma$, 3$\sigma$, and 4$\sigma$ of the WL map created with a $1\farcm9$ FWHM Gaussian kernel while dash-dotted silver contours are the same as in the top left panel. Bottom left panel: Optical image obtained from the Legacy Surveys / D. Lang (Perimeter Institute). Red stripes are saturation artifacts. The circles are the same as in the top left panel. The shadowed white-to-red filled contours are at [2,3,6,12,24,48]$\times\sigma$, with $\sigma=0.59$~mJy/beam, of the low-resolution uv-subtracted LOFAR image (97$\arcsec\times100\arcsec$). Bottom right panel: zoom-in view of the \textit{XMM-Newton} image shown in the top right panel. Orange solid contours are [3,6,12,24]$\times\sigma$, with $\sigma=0.58$~mJy/beam, of the uv-subtracted LOFAR image with restoring beam of 68$\arcsec\times81\arcsec$. The dash-dotted yellow contour is the 2$\sigma$ level of the low-resolution LOFAR image. The two white circles are $R_{500}$ and $R_{200}$ of PSZ2\,G083.29-31.03 estimated by \citet{Jia08}.}
      \label{fig:X-optical}
    \end{figure*}

The WL mass map was created using a color--color selected background galaxy sample, following the map-making procedure outlined in \citet{Umetsu2009}. In the upper right panel of Fig.~\ref{fig:X-optical}, we compare the resulting mass map, smoothed with a $1\farcm9$ FWHM Gaussian kernel, with the \textit{XMM-Newton} image. The rms reconstruction error, estimated from the $B$-mode map, is $\sigma_\kappa \simeq 0.03$, consistent with random shape noise. We also performed a mass reconstruction with a $3\farcm3$ FWHM Gaussian kernel, to better match the size of these structures (see the upper left panel of Fig.~\ref{fig:X-optical}).

The WL analysis provides additional evidence for the existence of a triple system and the mass distribution aligns well with that derived from X-ray emission and galaxy distribution.

\section{Discussion}
\label{sec:discussion}

The multiwavelength analysis gave us valuable insights into the dynamical state of the PSZ2\,G083.29-31.03 galaxy cluster. The presence of the central radio halo and the irregular X-ray morphology testify that PSZ2\,G083.29-31.03 is a merging galaxy cluster. The mass comparison between independent methods confirms that the node is very massive, but differences hint at an irregular and dynamically active system. The WL, SZ, or optical-based mass estimates differ to some degree, even though they are compatible within the statistical uncertainties and they are lower than the X-ray estimate. This might suggest that the gas temperature is temporarily boosted by the ongoing merger. 

The elongation toward the northwest of both X-ray and radio emission suggests that a component of the merger is along this direction. Beyond this, the presence of two structures is detected by X-rays, WL, and photometric analysis. The current data do not definitively clarify whether these smaller systems are the conclusive part of cosmic-web filaments channeling mass toward the central cluster, as claimed for example for Abell~2744 \citep{Eckert15Nat}, or if they are low-mass satellites located outside the outer regions of the cluster itself.

Considering the available data, we take into account the following potential interpretations for the detected large-scale radio emission: (i) associated with radio galaxies, (ii) radio relics, or (iii) radio halos. If the two sub-structures are indeed accreting into the central one, there is also the possibility that the diffuse emission originates from the dynamically active regions forming a massive node of the cosmic web (iv).

(i) The alignment of the two ‘‘ears’’ with the identified sub-structures is clear, which rules out the possibility of a coincidental alignment of radio emission associated with AGN activity. However, we note that a minor contribution from not fully subtracted sources to the flux density of the newly discovered emission cannot be excluded (see Appendix~\ref{app}).

(ii) Typical radio relics are elongated in the direction perpendicular to the merger axis that generates the central radio halo, while in this case, the merger axis of the central cluster is along the opposite direction. The two candidate relics would be located between the two sub-structures and the main cluster, where the radio emission appears elongated. This would suggest the presence of compression rather than shocks, but deeper X-ray observations, as well as multi-frequency radio data, are needed to firmly discard this hypothesis.

(iii) The mass of the two sub-structures is low (see Table~\ref{tab:clusters}). There is a scaling relation between the radio power of halos and the mass of the system \citep[e.g.,][]{Cuciti23} which indicates that in low-mass galaxy clusters, the amount of energy conveyed into non-thermal components is small.  A radio halo has been previously detected in a low-mass system ($M_{500}<2\times10^{14} M_{\odot}$), although at very low redshift and with low-resolution observations \citep{Botteon21}, and in this case its power was consistent with the extrapolation of the power-mass relation. The two ‘‘ears’’ have instead a power $P_{144{\rm MHz}}\sim1\times10^{25}(1+z)^{(\alpha-1)}$~W/Hz (where $\alpha$ is the unknown spectral index of the radio emission), which is one order of magnitude higher than that expected from the correlation. However, due to the large uncertainties in the mass of the systems, it is not possible to rule out the hypothesis that they are two powerful radio halos observed coincidentally aligned in projection. The high radio power could be also due to the mixture of different kinds of diffuse radio emissions.

(iv) The newly detected sources could finally represent the tip of the iceberg of the radio emission originating from the energy released by the mass accretion processes around a particularly massive node of the cosmic web. To understand if this system is unique we considered all the PSZ2 clusters in the Legacy Survey area (676). For each of them, we found the galaxy clusters and groups in a projected comoving radius of 5 Mpc and within a separation of $\Delta z = \pm 0.02 (1+z)$. In terms of the number of neighbors, PSZ2\,G083.29-31.03 is not peculiar: $\sim37\%$ of PSZ2 clusters have an equal or greater number of neighboring clusters and groups. However, if we limit the computation to neighbors with $M_{500}\geq10^{14} M_{\odot}$, PSZ2\,G083.29-31.03 is at the 90th percentile. When we consider the total mass of the neighbors, we again find that only $\sim10\%$ has an equal or greater mass in the surrounding systems.

The average surface brightness of the two ‘‘ears’’ (see Sec.~\ref{sec:radio}) is lower than 0.5~$\mu$Jy/arcsec$^2$ at 144 MHz, similarly to other systems where the diffuse radio emission extends outside $R_{500}$ and is sometimes referred to as mega-halos \citep[e.g][]{Shweta20,Rajpurohit21,Cuciti22,Botteon22b,Bruno23}. A similar value also characterizes the so-called bridges that connect nearby galaxy clusters \citep{Govoni19,Botteon20c,Pignataro24}. However, the average surface brightness obtained from stacked filaments between galaxy clusters is two orders of magnitude lower at the same frequency \citep{Vernstrom21}. If the detected emission results from energy released by accretion processes, deeper observations could potentially reveal its connection to the cosmic web.

With current data, the classification remains uncertain. However, considering the proximity of the three systems, multiple indicators of large-scale merger activity, and the radio power of the two "ears", we favor the last proposed scenario.

\section{Conclusions}
\label{sec:conclusions}

In this letter, we provide a comprehensive, radio, X-ray, optical, and weak lensing analysis of the PSZ2\,G083.29-31.03 galaxy cluster and its surroundings. In this system, we detected an extension to the previously known radio halo. The total extent of the 144 MHz radio emission reaches a projected linear extent of 5 Mpc. This emission is aligned with the X-ray emission detected by \textit{XMM-Newton} and the mass distribution derived from galaxies and weak lensing, which identified two low-mass systems at a redshift consistent with that of the main cluster, within the uncertainties. The detection was driven by the application of the Radio U-Net network \citep{Stuardi24} to the LoTSS survey. 

If the three structures are connected to the same massive node of the cosmic web, the newly detected emission would reach $R_{200}$. Emission on such large scales has been observed in an increasing number of massive ($M_{500}>5\times10^{14}M_\odot$) galaxy clusters; however, this would be the first case of large-scale radio emission linked to merging sub-structures at a relatively high redshift ($z=0.41$). Conversely, if the three structures are merely aligned in projection, the two faint sources could represent powerful radio halos hosted by low-mass clusters. In either scenario, this system warrants particular attention for follow-up studies.

Finally, this work demonstrates the potential of machine learning in exploiting vast datasets such as the one provided by current generation wide-area radio surveys such as LoTSS and EMU. These methodologies will be even more necessary with the approaching advent of the Square Kilometre Array\footnote{\url{https://www.skatelescope.org/}}.


\begin{acknowledgements}

We thank the Referee who stimulated the improvement of this manuscript.

This paper is supported by the Fondazione ICSC, Spoke 3 Astrophysics and Cosmos Observations. National Recovery and Resilience Plan (Piano Nazionale di Ripresa e Resilienza, PNRR) Project ID CN\_00000013 "Italian Research Center for High-Performance Computing, Big Data and Quantum Computing" funded by MUR Missione 4 Componente 2 Investimento 1.4: Potenziamento strutture di ricerca e creazione di "campioni nazionali di R\&S (M4C2-19)" - Next Generation EU (NGEU), and it's also supported by (Programma Operativo Nazionale, PON), ``Tematiche di Ricerca Green e dell'Innovazione". We acknowledge the CINECA award under the ISCRA initiative, for the availability of high performance computing resources and support. 

The HPC tests and benchmarks this work is based on, have been produced on the Leonardo Supercomputer at CINECA (Bologna, Italy) in the framework of the ISCRA programme IscrC\_RICK (project: HP10CDUNG6). 

The DESI Legacy Imaging Surveys consist of three individual and complementary projects: the Dark Energy Camera Legacy Survey (DECaLS), the Beijing-Arizona Sky Survey (BASS), and the Mayall z-band Legacy Survey (MzLS). DECaLS, BASS and MzLS together include data obtained, respectively, at the Blanco telescope, Cerro Tololo Inter-American Observatory, NSF’s NOIRLab; the Bok telescope, Steward Observatory, University of Arizona; and the Mayall telescope, Kitt Peak National Observatory, NOIRLab. NOIRLab is operated by the Association of Universities for Research in Astronomy (AURA) under a cooperative agreement with the National Science Foundation. Pipeline processing and analyses of the data were supported by NOIRLab and the Lawrence Berkeley National Laboratory (LBNL). Legacy Surveys also uses data products from the Near-Earth Object Wide-field Infrared Survey Explorer (NEOWISE), a project of the Jet Propulsion Laboratory/California Institute of Technology, funded by the National Aeronautics and Space Administration. Legacy Surveys was supported by: the Director, Office of Science, Office of High Energy Physics of the U.S. Department of Energy; the National Energy Research Scientific Computing Center, a DOE Office of Science User Facility; the U.S. National Science Foundation, Division of Astronomical Sciences; the National Astronomical Observatories of China, the Chinese Academy of Sciences and the Chinese National Natural Science Foundation. LBNL is managed by the Regents of the University of California under contract to the U.S. Department of Energy. The complete acknowledgments can be found at \url{https://www.legacysurvey.org/acknowledgment/}.

M.S. acknowledges the financial contribution from INAF Theory Grant 2023: Gravitational lensing detection of matter distribution at galaxy cluster boundaries and beyond (1.05.23.06.17) and from the contract Prin-MUR 2022 supported by Next Generation EU (M4.C2.1.1, n.20227RNLY3 {\it The concordance cosmological model: stress-tests with galaxy clusters}).

K.U. acknowledges support from the National Science and Technology Council of Taiwan (grant NSTC 112-2112-M-001-027-MY3) and the Academia Sinica Investigator Award (grant AS-IA-112-M04).

R.G. acknowledges the Agence Nationale de la Recherche (ANR Grant ‘AMALGAM’) and Centre National des Etudes Spatiales (CNES) for financial support. Part of the WL analysis benefited from the {\it infinity} and {\it morphomatic} computing facilities at IAP.

\end{acknowledgements}

%

\bibliographystyle{aa} 
\bibliography{my_bib,bib}

\begin{thebibliography}{48}
\expandafter\ifx\csname natexlab\endcsname\relax\def\natexlab#1{#1}\fi

\bibitem[{{Alegre} {et~al.}(2024){Alegre}, {Best}, {Sabater}, {Rottgering},
  {Hardcastle}, \& {Williams}}]{Alegre24}
{Alegre}, L., {Best}, P., {Sabater}, J., {et~al.} 2024, arXiv e-prints,
  arXiv:2405.18584

\bibitem[{{Amodeo} {et~al.}(2018){Amodeo}, {Mei}, {Stanford}, {Lawrence},
  {Bartlett}, {Stern}, {Chary}, {Shim}, {Marleau}, {Melin}, \&
  {Rodr{\'\i}guez-Gonz{\'a}lvez}}]{Amodeo18}
{Amodeo}, S., {Mei}, S., {Stanford}, S.~A., {et~al.} 2018, \apj, 853, 36

\bibitem[{{Applegate} {et~al.}(2014){Applegate}, {von der Linden}, {Kelly},
  {Allen}, {Allen}, {Burchat}, {Burke}, {Ebeling}, {Mantz}, \&
  {Morris}}]{wtg_III_14}
{Applegate}, D.~E., {von der Linden}, A., {Kelly}, P.~L., {et~al.} 2014,
  \mnras, 439, 48

\bibitem[{{Bade} {et~al.}(1998){Bade}, {Engels}, {Voges}, {Beckmann}, {Boller},
  {Cordis}, {Dahlem}, {Englhauser}, {Molthagen}, {Nass}, {Studt}, \&
  {Reimers}}]{Bade98}
{Bade}, N., {Engels}, D., {Voges}, W., {et~al.} 1998, \aaps, 127, 145

\bibitem[{{Beduzzi} {et~al.}(2023){Beduzzi}, {Vazza}, {Brunetti}, {Cuciti},
  {Wittor}, \& {Corsini}}]{Beduzzi23}
{Beduzzi}, L., {Vazza}, F., {Brunetti}, G., {et~al.} 2023, \aap, 678, L8

\bibitem[{{Botteon} {et~al.}(2021){Botteon}, {Cassano}, {van Weeren},
  {Shimwell}, {Bonafede}, {Brggen}, {Brunetti}, {Cuciti}, {Dallacasa}, {de
  Gasperin}, {Di Gennaro}, {Gastaldello}, {Hoang}, {Rossetti}, \&
  {R{\"o}ttgering}}]{Botteon21}
{Botteon}, A., {Cassano}, R., {van Weeren}, R.~J., {et~al.} 2021, \apjl, 914,
  L29

\bibitem[{{Botteon} {et~al.}(2022{\natexlab{a}}){Botteon}, {Shimwell},
  {Cassano}, {Cuciti}, {Zhang}, {Bruno}, {Camillini}, {Natale}, {Jones},
  {Gastaldello}, {Simionescu}, {Rossetti}, {Akamatsu}, {van Weeren},
  {Brunetti}, {Br{\"u}ggen}, {Groeneveld}, {Hoang}, {Hardcastle}, {Ignesti},
  {Di Gennaro}, {Bonafede}, {Drabent}, {R{\"o}ttgering}, {Hoeft}, \& {de
  Gasperin}}]{Botteon22}
{Botteon}, A., {Shimwell}, T.~W., {Cassano}, R., {et~al.} 2022{\natexlab{a}},
  \aap, 660, A78

\bibitem[{{Botteon} {et~al.}(2020){Botteon}, {van Weeren}, {Brunetti}, {de
  Gasperin}, {Intema}, {Osinga}, {Di Gennaro}, {Shimwell}, {Bonafede},
  {Br{\"u}ggen}, {Cassano}, {Cuciti}, {Dallacasa}, {Gastaldello}, {Mandal},
  {Rossetti}, \& {R{\"o}ttgering}}]{Botteon20c}
{Botteon}, A., {van Weeren}, R.~J., {Brunetti}, G., {et~al.} 2020, \mnras, 499,
  L11

\bibitem[{{Botteon} {et~al.}(2022{\natexlab{b}}){Botteon}, {van Weeren},
  {Brunetti}, {Vazza}, {Shimwell}, {Br{\"u}ggen}, {R{\"o}ttgering}, {de
  Gasperin}, {Akamatsu}, {Bonafede}, {Cassano}, {Cuciti}, {Dallacasa}, {Di
  Gennaro}, \& {Gastaldello}}]{Botteon22b}
{Botteon}, A., {van Weeren}, R.~J., {Brunetti}, G., {et~al.}
  2022{\natexlab{b}}, Science Advances, 8, eabq7623

\bibitem[{{Bruno} {et~al.}(2023){Bruno}, {Botteon}, {Shimwell}, {Cuciti}, {de
  Gasperin}, {Brunetti}, {Dallacasa}, {Gastaldello}, {Rossetti}, {van Weeren},
  {Venturi}, {Russo}, {Taffoni}, {Cassano}, {Biava}, {Lusetti}, {Bonafede},
  {Ghizzardi}, \& {De Grandi}}]{Bruno23}
{Bruno}, L., {Botteon}, A., {Shimwell}, T., {et~al.} 2023, \aap, 678, A133

\bibitem[{{CHEX-MATE Collaboration} {et~al.}(2021){CHEX-MATE Collaboration},
  {Arnaud}, {Ettori}, {Pratt}, {Rossetti}, {Eckert}, {Gastaldello}, {Gavazzi},
  {Kay}, {Lovisari}, {Maughan}, {Pointecouteau}, {Sereno}, {Bartalucci},
  {Bonafede}, {Bourdin}, {Cassano}, {Duffy}, {Iqbal}, {Maurogordato}, {Rasia},
  {Sayers}, {Andrade-Santos}, {Aussel}, {Barnes}, {Barrena}, {Borgani},
  {Burkutean}, {Clerc}, {Corasaniti}, {Cuillandre}, {De Grandi}, {De Petris},
  {Dolag}, {Donahue}, {Ferragamo}, {Gaspari}, {Ghizzardi}, {Gitti}, {Haines},
  {Jauzac}, {Johnston-Hollitt}, {Jones}, {K{\'e}ruzor{\'e}}, {Le Brun},
  {Mayet}, {Mazzotta}, {Melin}, {Molendi}, {Nonino}, {Okabe}, {Paltani},
  {Perotto}, {Pires}, {Radovich}, {Rubino-Martin}, {Salvati}, {Saro},
  {Sartoris}, {Schellenberger}, {Streblyanska}, {Tarr{\'\i}o}, {Tozzi},
  {Umetsu}, {van der Burg}, {Vazza}, {Venturi}, {Yepes}, \&
  {Zarattini}}]{CHEX-MATE21}
{CHEX-MATE Collaboration}, {Arnaud}, M., {Ettori}, S., {et~al.} 2021, \aap,
  650, A104

\bibitem[{{Cuciti} {et~al.}(2023){Cuciti}, {Cassano}, {Sereno}, {Brunetti},
  {Botteon}, {Shimwell}, {Bruno}, {Gastaldello}, {Rossetti}, {Zhang},
  {Simionescu}, {Br{\"u}ggen}, {van Weeren}, {Jones}, {Akamatsu}, {Bonafede},
  {De Gasperin}, {Di Gennaro}, {Pasini}, \& {R{\"o}ttgering}}]{Cuciti23}
{Cuciti}, V., {Cassano}, R., {Sereno}, M., {et~al.} 2023, \aap, 680, A30

\bibitem[{{Cuciti} {et~al.}(2022){Cuciti}, {de Gasperin}, {Br{\"u}ggen},
  {Vazza}, {Brunetti}, {Shimwell}, {Edler}, {van Weeren}, {Botteon}, {Cassano},
  {Di Gennaro}, {Gastaldello}, {Drabent}, {R{\"o}ttgering}, \&
  {Tasse}}]{Cuciti22}
{Cuciti}, V., {de Gasperin}, F., {Br{\"u}ggen}, M., {et~al.} 2022, \nat, 609,
  911

\bibitem[{{Dey} {et~al.}(2019){Dey}, {Schlegel}, {Lang}, {Blum}, {Burleigh},
  {Fan}, {Findlay}, {Finkbeiner}, {Herrera}, {Juneau}, {Landriau}, {Levi},
  {McGreer}, {Meisner}, {Myers}, {Moustakas}, {Nugent}, {Patej}, {Schlafly},
  {Walker}, {Valdes}, {Weaver}, {Y{\`e}che}, {Zou}, {Zhou}, {Abareshi},
  {Abbott}, {Abolfathi}, {Aguilera}, {Alam}, {Allen}, {Alvarez}, {Annis},
  {Ansarinejad}, {Aubert}, {Beechert}, {Bell}, {BenZvi}, {Beutler}, {Bielby},
  {Bolton}, {Brice{\~n}o}, {Buckley-Geer}, {Butler}, {Calamida}, {Carlberg},
  {Carter}, {Casas}, {Castander}, {Choi}, {Comparat}, {Cukanovaite}, {Delubac},
  {DeVries}, {Dey}, {Dhungana}, {Dickinson}, {Ding}, {Donaldson}, {Duan},
  {Duckworth}, {Eftekharzadeh}, {Eisenstein}, {Etourneau}, {Fagrelius},
  {Farihi}, {Fitzpatrick}, {Font-Ribera}, {Fulmer}, {G{\"a}nsicke},
  {Gaztanaga}, {George}, {Gerdes}, {Gontcho}, {Gorgoni}, {Green}, {Guy},
  {Harmer}, {Hernandez}, {Honscheid}, {Huang}, {James}, {Jannuzi}, {Jiang},
  {Joyce}, {Karcher}, {Karkar}, {Kehoe}, {Kneib}, {Kueter-Young}, {Lan},
  {Lauer}, {Le Guillou}, {Le Van Suu}, {Lee}, {Lesser}, {Perreault Levasseur},
  {Li}, {Mann}, {Marshall}, {Mart{\'\i}nez-V{\'a}zquez}, {Martini}, {du Mas des
  Bourboux}, {McManus}, {Meier}, {M{\'e}nard}, {Metcalfe},
  {Mu{\~n}oz-Guti{\'e}rrez}, {Najita}, {Napier}, {Narayan}, {Newman}, {Nie},
  {Nord}, {Norman}, {Olsen}, {Paat}, {Palanque-Delabrouille}, {Peng},
  {Poppett}, {Poremba}, {Prakash}, {Rabinowitz}, {Raichoor}, {Rezaie},
  {Robertson}, {Roe}, {Ross}, {Ross}, {Rudnick}, {Safonova}, {Saha},
  {S{\'a}nchez}, {Savary}, {Schweiker}, {Scott}, {Seo}, {Shan}, {Silva},
  {Slepian}, {Soto}, {Sprayberry}, {Staten}, {Stillman}, {Stupak}, {Summers},
  {Sien Tie}, {Tirado}, {Vargas-Maga{\~n}a}, {Vivas}, {Wechsler}, {Williams},
  {Yang}, {Yang}, {Yapici}, {Zaritsky}, {Zenteno}, {Zhang}, {Zhang}, {Zhou}, \&
  {Zhou}}]{DESI_dey+al19}
{Dey}, A., {Schlegel}, D.~J., {Lang}, D., {et~al.} 2019, \aj, 157, 168

\bibitem[{{Eckert} {et~al.}(2015){Eckert}, {Jauzac}, {Shan}, {Kneib}, {Erben},
  {Israel}, {Jullo}, {Klein}, {Massey}, {Richard}, \& {Tchernin}}]{Eckert15Nat}
{Eckert}, D., {Jauzac}, M., {Shan}, H., {et~al.} 2015, \nat, 528, 105

\bibitem[{{Fo{\"e}x} {et~al.}(2012){Fo{\"e}x}, {Soucail}, {Pointecouteau},
  {Arnaud}, {Limousin}, \& {Pratt}}]{foe+al12}
{Fo{\"e}x}, G., {Soucail}, G., {Pointecouteau}, E., {et~al.} 2012, \aap, 546,
  A106

\bibitem[{{Gheller} {et~al.}(2018){Gheller}, {Vazza}, \&
  {Bonafede}}]{Gheller18}
{Gheller}, C., {Vazza}, F., \& {Bonafede}, A. 2018, \mnras, 480, 3749

\bibitem[{{Giovannini} {et~al.}(2020){Giovannini}, {Cau}, {Bonafede},
  {Ebeling}, {Feretti}, {Girardi}, {Gitti}, {Govoni}, {Ignesti}, {Murgia},
  {Taylor}, \& {Vacca}}]{Giovannini20}
{Giovannini}, G., {Cau}, M., {Bonafede}, A., {et~al.} 2020, \aap, 640, A108

\bibitem[{{Govoni} {et~al.}(2019){Govoni}, {Orr{\`u}}, {Bonafede}, {Iacobelli},
  {Paladino}, {Vazza}, {Murgia}, {Vacca}, {Giovannini}, {Feretti}, {Loi},
  {Bernardi}, {Ferrari}, {Pizzo}, {Gheller}, {Manti}, {Br{\"u}ggen},
  {Brunetti}, {Cassano}, {de Gasperin}, {En{\ss}lin}, {Hoeft}, {Horellou},
  {Junklewitz}, {R{\"o}ttgering}, {Scaife}, {Shimwell}, {van Weeren}, \&
  {Wise}}]{Govoni19}
{Govoni}, F., {Orr{\`u}}, E., {Bonafede}, A., {et~al.} 2019, Science, 364, 981

\bibitem[{{Gupta} {et~al.}(2024){Gupta}, {Norris}, {Hayder}, {Huynh},
  {Petersson}, {Rosalind Wang}, {Hopkins}, {Andernach}, {Gordon}, {Riggi},
  {Yew}, {Crawford}, {Koribalski}, {Filipovi{\'c}}, {Kapi{\'n}ska}, {Shabala},
  {Vernstrom}, \& {Marvil}}]{Gupta24}
{Gupta}, N., {Norris}, R.~P., {Hayder}, Z., {et~al.} 2024, \pasa, 41, e027

\bibitem[{{Jia} {et~al.}(2008){Jia}, {B{\"o}hringer}, {Pointecouteau}, {Chen},
  \& {Zhang}}]{Jia08}
{Jia}, S.~M., {B{\"o}hringer}, H., {Pointecouteau}, E., {Chen}, Y., \& {Zhang},
  Y.~Y. 2008, \aap, 489, 1

\bibitem[{{Lastufka} {et~al.}(2024){Lastufka}, {Bait}, {Taran}, {Drozdova},
  {Kinakh}, {Piras}, {Audard}, {Dessauges-Zavadsky}, {Holotyak}, {Schaerer}, \&
  {Voloshynovskiy}}]{Lastufka24}
{Lastufka}, E., {Bait}, O., {Taran}, O., {et~al.} 2024, arXiv e-prints,
  arXiv:2408.06147

\bibitem[{{Mandelbaum} {et~al.}(2015){Mandelbaum}, {Rowe}, {Armstrong}, {Bard},
  {Bertin}, {Bosch}, {Boutigny}, {Courbin}, {Dawson}, {Donnarumma}, {Fenech
  Conti}, {Gavazzi}, {Gentile}, {Gill}, {Hogg}, {Huff}, {Jee}, {Kacprzak},
  {Kilbinger}, {Kuntzer}, {Lang}, {Luo}, {March}, {Marshall}, {Meyers},
  {Miller}, {Miyatake}, {Nakajima}, {Ngol{\'e} Mboula}, {Nurbaeva}, {Okura},
  {Paulin-Henriksson}, {Rhodes}, {Schneider}, {Shan}, {Sheldon}, {Simet},
  {Starck}, {Sureau}, {Tewes}, {Zarb Adami}, {Zhang}, \&
  {Zuntz}}]{Mandelbaum14}
{Mandelbaum}, R., {Rowe}, B., {Armstrong}, R., {et~al.} 2015, \mnras, 450, 2963

\bibitem[{{Mostert} {et~al.}(2021){Mostert}, {Duncan}, {R{\"o}ttgering},
  {Polsterer}, {Best}, {Brienza}, {Br{\"u}ggen}, {Hardcastle}, {Jurlin},
  {Mingo}, {Morganti}, {Shimwell}, {Smith}, \& {Williams}}]{Mostert21}
{Mostert}, R. I.~J., {Duncan}, K.~J., {R{\"o}ttgering}, H. J.~A., {et~al.}
  2021, \aap, 645, A89

\bibitem[{{Nishiwaki} {et~al.}(2024){Nishiwaki}, {Brunetti}, {Vazza}, \&
  {Gheller}}]{Nishiwaki24}
{Nishiwaki}, K., {Brunetti}, G., {Vazza}, F., \& {Gheller}, C. 2024, \apj, 961,
  15

\bibitem[{{Norris} {et~al.}(2011){Norris}, {Hopkins}, {Afonso}, {Brown},
  {Condon}, {Dunne}, {Feain}, {Hollow}, {Jarvis}, {Johnston-Hollitt}, {Lenc},
  {Middelberg}, {Padovani}, {Prandoni}, {Rudnick}, {Seymour}, {Umana},
  {Andernach}, {Alexander}, {Appleton}, {Bacon}, {Banfield}, {Becker}, {Brown},
  {Ciliegi}, {Jackson}, {Eales}, {Edge}, {Gaensler}, {Giovannini}, {Hales},
  {Hancock}, {Huynh}, {Ibar}, {Ivison}, {Kennicutt}, {Kimball}, {Koekemoer},
  {Koribalski}, {L{\'o}pez-S{\'a}nchez}, {Mao}, {Murphy}, {Messias},
  {Pimbblet}, {Raccanelli}, {Randall}, {Reiprich}, {Roseboom},
  {R{\"o}ttgering}, {Saikia}, {Sharp}, {Slee}, {Smail}, {Thompson}, {Urquhart},
  {Wall}, \& {Zhao}}]{Norris11}
{Norris}, R.~P., {Hopkins}, A.~M., {Afonso}, J., {et~al.} 2011, \pasa, 28, 215

\bibitem[{Offringa {et~al.}(2014)Offringa, McKinley, Hurley-Walker,
  {et~al.}}]{Offringa14}
Offringa, A.~R., McKinley, B., Hurley-Walker, {et~al.} 2014, MNRAS, 444, 606

\bibitem[{Offringa \& Smirnov(2017)}]{Offringa17}
Offringa, A.~R. \& Smirnov, O. 2017, MNRAS, 471, 301

\bibitem[{{Pignataro} {et~al.}(2024){Pignataro}, {Bonafede}, {Bernardi},
  {Balboni}, {Vazza}, {van Weeren}, {Ubertosi}, {Cassano}, {Brunetti},
  {Botteon}, {Venturi}, {Akamatsu}, {Drabent}, \& {Hoeft}}]{Pignataro24}
{Pignataro}, G.~V., {Bonafede}, A., {Bernardi}, G., {et~al.} 2024, arXiv
  e-prints, arXiv:2409.15412

\bibitem[{{Planck Collaboration} {et~al.}(2016){Planck Collaboration}, {Ade},
  {Aghanim}, {Arnaud}, {Ashdown}, {Aumont}, {Baccigalupi}, {Banday},
  {Barreiro}, {Barrena}, {Bartlett}, {Bartolo}, {Battaner}, {Battye},
  {Benabed}, {Beno{\^\i}t}, {Benoit-L{\'e}vy}, {Bernard}, {Bersanelli},
  {Bielewicz}, {Bikmaev}, {B{\"o}hringer}, {Bonaldi}, {Bonavera}, {Bond},
  {Borrill}, {Bouchet}, {Bucher}, {Burenin}, {Burigana}, {Butler}, {Calabrese},
  {Cardoso}, {Carvalho}, {Catalano}, {Challinor}, {Chamballu}, {Chary},
  {Chiang}, {Chon}, {Christensen}, {Clements}, {Colombi}, {Colombo}, {Combet},
  {Comis}, {Couchot}, {Coulais}, {Crill}, {Curto}, {Cuttaia}, {Dahle},
  {Danese}, {Davies}, {Davis}, {de Bernardis}, {de Rosa}, {de Zotti},
  {Delabrouille}, {D{\'e}sert}, {Dickinson}, {Diego}, {Dolag}, {Dole},
  {Donzelli}, {Dor{\'e}}, {Douspis}, {Ducout}, {Dupac}, {Efstathiou},
  {Eisenhardt}, {Elsner}, {En{\ss}lin}, {Eriksen}, {Falgarone}, {Fergusson},
  {Feroz}, {Ferragamo}, {Finelli}, {Forni}, {Frailis}, {Fraisse}, {Franceschi},
  {Frejsel}, {Galeotta}, {Galli}, {Ganga}, {G{\'e}nova-Santos}, {Giard},
  {Giraud-H{\'e}raud}, {Gjerl{\o}w}, {Gonz{\'a}lez-Nuevo}, {G{\'o}rski},
  {Grainge}, {Gratton}, {Gregorio}, {Gruppuso}, {Gudmundsson}, {Hansen},
  {Hanson}, {Harrison}, {Hempel}, {Henrot- Versill{\'e}},
  {Hern{\'a}ndez-Monteagudo}, {Herranz}, {Hildebrandt}, {Hivon}, {Hobson},
  {Holmes}, {Hornstrup}, {Hovest}, {Huffenberger}, {Hurier}, {Jaffe}, {Jaffe},
  {Jin}, {Jones}, {Juvela}, {Keih{\"a}nen}, {Keskitalo}, {Khamitov}, {Kisner},
  {Kneissl}, {Knoche}, {Kunz}, {Kurki-Suonio}, {Lagache}, {Lamarre}, {Lasenby},
  {Lattanzi}, {Lawrence}, {Leonardi}, {Lesgourgues}, {Levrier}, {Liguori},
  {Lilje}, {Linden-V{\o}rnle}, {L{\'o}pez- Caniego}, {Lubin},
  {Mac{\'\i}as-P{\'e}rez}, {Maggio}, {Maino}, {Mak}, {Mandolesi}, {Mangilli},
  {Martin}, {Mart{\'\i}nez-Gonz{\'a}lez}, {Masi}, {Matarrese}, {Mazzotta},
  {McGehee}, {Mei}, {Melchiorri}, {Melin}, {Mendes}, {Mennella}, {Migliaccio},
  {Mitra}, {Miville-Desch{\^e}nes}, {Moneti}, {Montier}, {Morgante},
  {Mortlock}, {Moss}, {Munshi}, {Murphy}, {Naselsky}, {Nastasi}, {Nati},
  {Natoli}, {Netterfield}, {N{\o}rgaard-Nielsen}, {Noviello}, {Novikov},
  {Novikov}, {Olamaie}, {Oxborrow}, {Paci}, {Pagano}, {Pajot}, {Paoletti},
  {Pasian}, {Patanchon}, {Pearson}, {Perdereau}, {Perotto}, {Perrott},
  {Perrotta}, {Pettorino}, {Piacentini}, {Piat}, {Pierpaoli}, {Pietrobon},
  {Plaszczynski}, {Pointecouteau}, {Polenta}, {Pratt}, {Pr{\'e}zeau}, {Prunet},
  {Puget}, {Rachen}, {Reach}, {Rebolo}, {Reinecke}, {Remazeilles}, {Renault},
  {Renzi}, {Ristorcelli}, {Rocha}, {Rosset}, {Rossetti}, {Roudier}, {Rozo},
  {Rubi{\~n}o-Mart{\'\i}n}, {Rumsey}, {Rusholme}, {Rykoff}, {Sandri}, {Santos},
  {Saunders}, {Savelainen}, {Savini}, {Schammel}, {Scott}, {Seiffert},
  {Shellard}, {Shimwell}, {Spencer}, {Stanford}, {Stern}, {Stolyarov},
  {Stompor}, {Streblyanska}, {Sudiwala}, {Sunyaev}, {Sutton}, {Suur-Uski},
  {Sygnet}, {Tauber}, {Terenzi}, {Toffolatti}, {Tomasi}, {Tramonte},
  {Tristram}, {Tucci}, {Tuovinen}, {Umana}, {Valenziano}, {Valiviita}, {Van
  Tent}, {Vielva}, {Villa}, {Wade}, {Wandelt}, {Wehus}, {White}, {Wright},
  {Yvon}, {Zacchei}, \& {Zonca}}]{Planck16b}
{Planck Collaboration}, {Ade}, P.~A.~R., {Aghanim}, N., {et~al.} 2016, {\aap},
  594, A27

\bibitem[{{Pointecouteau} {et~al.}(2002){Pointecouteau}, {Hattori}, {Neumann},
  {Komatsu}, {Matsuo}, {Kuno}, \& {B{\"o}hringer}}]{Pointecouteau02}
{Pointecouteau}, E., {Hattori}, M., {Neumann}, D., {et~al.} 2002, \aap, 387, 56

\bibitem[{{Rajpurohit} {et~al.}(2021){Rajpurohit}, {Vazza}, {van Weeren},
  {Hoeft}, {Brienza}, {Bonnassieux}, {Riseley}, {Brunetti}, {Bonafede},
  {Br{\"u}ggen}, {Formann}, {Rajpurohit}, {R{\"o}ttgering}, {Drabent},
  {Dom{\'\i}nguez-Fern{\'a}ndez}, {Wittor}, \& {Andrade-Santos}}]{Rajpurohit21}
{Rajpurohit}, K., {Vazza}, F., {van Weeren}, R.~J., {et~al.} 2021, \aap, 654,
  A41

\bibitem[{{Sereno}(2015)}]{ser15_comalit_III}
{Sereno}, M. 2015, \mnras, 450, 3665

\bibitem[{{Sereno} {et~al.}(2025){Sereno}, {Maurogordato}, {Cappi}, {Barrena},
  {Benoist}, {Haines}, {Radovich}, {Nonino}, {Ettori}, {Ferragamo}, {Gavazzi},
  {Huot}, {Pizzuti}, {Pratt}, {Streblyanska}, {Zarattini}, {Castignani},
  {Eckert}, {Gastaldello}, {Kay}, {Lovisari}, {Maughan}, {Pointecouteau},
  {Rasia}, {Rossetti}, \& {Sayers}}]{Sereno25}
{Sereno}, M., {Maurogordato}, S., {Cappi}, A., {et~al.} 2025, \aap, 693, A2

\bibitem[{{Shimwell} {et~al.}(2022){Shimwell}, {Hardcastle}, {Tasse}, {Best},
  {R{\"o}ttgering}, {Williams}, {Botteon}, {Drabent}, {Mechev}, {Shulevski},
  {van Weeren}, {Bester}, {Br{\"u}ggen}, {Brunetti}, {Callingham}, {Chy{\.z}y},
  {Conway}, {Dijkema}, {Duncan}, {de Gasperin}, {Hale}, {Haverkorn}, {Hugo},
  {Jackson}, {Mevius}, {Miley}, {Morabito}, {Morganti}, {Offringa}, {Oonk},
  {Rafferty}, {Sabater}, {Smith}, {Schwarz}, {Smirnov}, {O'Sullivan},
  {Vedantham}, {White}, {Albert}, {Alegre}, {Asabere}, {Bacon}, {Bonafede},
  {Bonnassieux}, {Brienza}, {Bilicki}, {Bonato}, {Calistro Rivera}, {Cassano},
  {Cochrane}, {Croston}, {Cuciti}, {Dallacasa}, {Danezi}, {Dettmar}, {Di
  Gennaro}, {Edler}, {En{\ss}lin}, {Emig}, {Franzen}, {Garc{\'\i}a-Vergara},
  {Grange}, {G{\"u}rkan}, {Hajduk}, {Heald}, {Heesen}, {Hoang}, {Hoeft},
  {Horellou}, {Iacobelli}, {Jamrozy}, {Jeli{\'c}}, {Kondapally}, {Kukreti},
  {Kunert-Bajraszewska}, {Magliocchetti}, {Mahatma}, {Ma{\l}ek}, {Mandal},
  {Massaro}, {Meyer-Zhao}, {Mingo}, {Mostert}, {Nair}, {Nakoneczny},
  {Nikiel-Wroczy{\'n}ski}, {Orr{\'u}}, {Pajdosz-{\'S}mierciak}, {Pasini},
  {Prandoni}, {van Piggelen}, {Rajpurohit}, {Retana-Montenegro}, {Riseley},
  {Rowlinson}, {Saxena}, {Schrijvers}, {Sweijen}, {Siewert}, {Timmerman},
  {Vaccari}, {Vink}, {West}, {Wo{\l}owska}, {Zhang}, \& {Zheng}}]{Shimwell22}
{Shimwell}, T.~W., {Hardcastle}, M.~J., {Tasse}, C., {et~al.} 2022, \aap, 659,
  A1

\bibitem[{{Shimwell} {et~al.}(2017){Shimwell}, {R{\"o}ttgering}, {Best},
  {Williams}, {Dijkema}, {de Gasperin}, {Hardcastle}, {Heald}, {Hoang},
  {Horneffer}, {Intema}, {Mahony}, {Mandal}, {Mechev}, {Morabito}, {Oonk},
  {Rafferty}, {Retana-Montenegro}, {Sabater}, {Tasse}, {van Weeren},
  {Br{\"u}ggen}, {Brunetti}, {Chy{\.z}y}, {Conway}, {Haverkorn}, {Jackson},
  {Jarvis}, {McKean}, {Miley}, {Morganti}, {White}, {Wise}, {van Bemmel},
  {Beck}, {Brienza}, {Bonafede}, {Calistro Rivera}, {Cassano}, {Clarke},
  {Cseh}, {Deller}, {Drabent}, {van Driel}, {Engels}, {Falcke}, {Ferrari},
  {Fr{\"o}hlich}, {Garrett}, {Harwood}, {Heesen}, {Hoeft}, {Horellou},
  {Israel}, {Kapi{\'n}ska}, {Kunert-Bajraszewska}, {McKay}, {Mohan},
  {Orr{\'u}}, {Pizzo}, {Prandoni}, {Schwarz}, {Shulevski}, {Sipior}, {Smith},
  {Sridhar}, {Steinmetz}, {Stroe}, {Varenius}, {van der Werf}, {Zensus}, \&
  {Zwart}}]{Shimwell17}
{Shimwell}, T.~W., {R{\"o}ttgering}, H.~J.~A., {Best}, P.~N., {et~al.} 2017,
  {\aap}, 598, A104

\bibitem[{{Shimwell} {et~al.}(2019){Shimwell}, {Tasse}, {Hardcastle}, {Mechev},
  {Williams}, {Best}, {R{\"o}ttgering}, {Callingham}, {Dijkema}, {de Gasperin},
  {Hoang}, {Hugo}, {Mirmont}, {Oonk}, {Prandoni}, {Rafferty}, {Sabater},
  {Smirnov}, {van Weeren}, {White}, {Atemkeng}, {Bester}, {Bonnassieux},
  {Br{\"u}ggen}, {Brunetti}, {Chy{\.z}y}, {Cochrane}, {Conway}, {Croston},
  {Danezi}, {Duncan}, {Haverkorn}, {Heald}, {Iacobelli}, {Intema}, {Jackson},
  {Jamrozy}, {Jarvis}, {Lakhoo}, {Mevius}, {Miley}, {Morabito}, {Morganti},
  {Nisbet}, {Orr{\'u}}, {Perkins}, {Pizzo}, {Schrijvers}, {Smith}, {Vermeulen},
  {Wise}, {Alegre}, {Bacon}, {van Bemmel}, {Beswick}, {Bonafede}, {Botteon},
  {Bourke}, {Brienza}, {Calistro Rivera}, {Cassano}, {Clarke}, {Conselice},
  {Dettmar}, {Drabent}, {Dumba}, {Emig}, {En{\ss}lin}, {Ferrari}, {Garrett},
  {G{\'e}nova-Santos}, {Goyal}, {G{\"u}rkan}, {Hale}, {Harwood}, {Heesen},
  {Hoeft}, {Horellou}, {Jackson}, {Kokotanekov}, {Kondapally},
  {Kunert-Bajraszewska}, {Mahatma}, {Mahony}, {Mandal}, {McKean}, {Merloni},
  {Mingo}, {Miskolczi}, {Mooney}, {Nikiel-Wroczy{\'n}ski}, {O'Sullivan},
  {Quinn}, {Reich}, {Roskowi{\'n}ski}, {Rowlinson}, {Savini}, {Saxena},
  {Schwarz}, {Shulevski}, {Sridhar}, {Stacey}, {Urquhart}, {van der Wiel},
  {Varenius}, {Webster}, \& {Wilber}}]{Shimwell19}
{Shimwell}, T.~W., {Tasse}, C., {Hardcastle}, M.~J., {et~al.} 2019, {\aap},
  622, A1

\bibitem[{{Shweta} {et~al.}(2020){Shweta}, {Athreya}, \& {Sekhar}}]{Shweta20}
{Shweta}, A., {Athreya}, R., \& {Sekhar}, S. 2020, \apj, 897, 115

\bibitem[{{Slijepcevic} {et~al.}(2024){Slijepcevic}, {Scaife}, {Walmsley},
  {Bowles}, {Wong}, {Shabala}, \& {White}}]{Slijepcevic24}
{Slijepcevic}, I.~V., {Scaife}, A. M.~M., {Walmsley}, M., {et~al.} 2024, RAS
  Techniques and Instruments, 3, 19

\bibitem[{{Snowden} {et~al.}(2008){Snowden}, {Mushotzky}, {Kuntz}, \&
  {Davis}}]{Snowden08}
{Snowden}, S.~L., {Mushotzky}, R.~F., {Kuntz}, K.~D., \& {Davis}, D.~S. 2008,
  {\aap}, 478, 615

\bibitem[{{Stuardi} {et~al.}(2024){Stuardi}, {Gheller}, {Vazza}, \&
  {Botteon}}]{Stuardi24}
{Stuardi}, C., {Gheller}, C., {Vazza}, F., \& {Botteon}, A. 2024, \mnras, 533,
  3194

\bibitem[{{Umetsu}(2020)}]{Umetsu2020}
{Umetsu}, K. 2020, \aapr, 28, 7

\bibitem[{{Umetsu} {et~al.}(2009){Umetsu}, {Birkinshaw}, {Liu}, {Wu},
  {Medezinski}, {Broadhurst}, {Lemze}, {Zitrin}, {Ho}, {Huang}, {Koch}, {Liao},
  {Lin}, {Molnar}, {Nishioka}, {Wang}, {Altamirano}, {Chang}, {Chang}, {Chang},
  {Chen}, {Han}, {Huang}, {Hwang}, {Jiang}, {Kesteven}, {Kubo}, {Li},
  {Martin-Cocher}, {Oshiro}, {Raffin}, {Wei}, \& {Wilson}}]{Umetsu2009}
{Umetsu}, K., {Birkinshaw}, M., {Liu}, G.-C., {et~al.} 2009, \apj, 694, 1643

\bibitem[{{van Haarlem} {et~al.}(2013){van Haarlem}, {Wise}, {Gunst}, {Heald},
  {McKean}, {Hessels}, {de Bruyn}, {Nijboer}, {Swinbank}, {Fallows},
  {Brentjens}, {Nelles}, {Beck}, {Falcke}, {Fender}, {H{\"o}randel},
  {Koopmans}, {Mann}, {Miley}, {R{\"o}ttgering}, {Stappers}, {Wijers},
  {Zaroubi}, {van den Akker}, {Alexov}, {Anderson}, {Anderson}, {van Ardenne},
  {Arts}, {Asgekar}, {Avruch}, {Batejat}, {B{\"a}hren}, {Bell}, {Bell}, {van
  Bemmel}, {Bennema}, {Bentum}, {Bernardi}, {Best}, {B{\^\i}rzan}, {Bonafede},
  {Boonstra}, {Braun}, {Bregman}, {Breitling}, {van de Brink}, {Broderick},
  {Broekema}, {Brouw}, {Br{\"u}ggen}, {Butcher}, {van Cappellen}, {Ciardi},
  {Coenen}, {Conway}, {Coolen}, {Corstanje}, {Damstra}, {Davies}, {Deller},
  {Dettmar}, {van Diepen}, {Dijkstra}, {Donker}, {Doorduin}, {Dromer}, {Drost},
  {van Duin}, {Eisl{\"o}ffel}, {van Enst}, {Ferrari}, {Frieswijk}, {Gankema},
  {Garrett}, {de Gasperin}, {Gerbers}, {de Geus}, {Grie{\ss}meier}, {Grit},
  {Gruppen}, {Hamaker}, {Hassall}, {Hoeft}, {Holties}, {Horneffer}, {van der
  Horst}, {van Houwelingen}, {Huijgen}, {Iacobelli}, {Intema}, {Jackson},
  {Jelic}, {de Jong}, {Juette}, {Kant}, {Karastergiou}, {Koers}, {Kollen},
  {Kondratiev}, {Kooistra}, {Koopman}, {Koster}, {Kuniyoshi}, {Kramer},
  {Kuper}, {Lambropoulos}, {Law}, {van Leeuwen}, {Lemaitre}, {Loose}, {Maat},
  {Macario}, {Markoff}, {Masters}, {McFadden}, {McKay-Bukowski}, {Meijering},
  {Meulman}, {Mevius}, {Middelberg}, {Millenaar}, {Miller-Jones}, {Mohan},
  {Mol}, {Morawietz}, {Morganti}, {Mulcahy}, {Mulder}, {Munk}, {Nieuwenhuis},
  {van Nieuwpoort}, {Noordam}, {Norden}, {Noutsos}, {Offringa}, {Olofsson},
  {Omar}, {Orr{\'u}}, {Overeem}, {Paas}, {Pand ey-Pommier}, {Pandey}, {Pizzo},
  {Polatidis}, {Rafferty}, {Rawlings}, {Reich}, {de Reijer}, {Reitsma},
  {Renting}, {Riemers}, {Rol}, {Romein}, {Roosjen}, {Ruiter}, {Scaife}, {van
  der Schaaf}, {Scheers}, {Schellart}, {Schoenmakers}, {Schoonderbeek},
  {Serylak}, {Shulevski}, {Sluman}, {Smirnov}, {Sobey}, {Spreeuw}, {Steinmetz},
  {Sterks}, {Stiepel}, {Stuurwold}, {Tagger}, {Tang}, {Tasse}, {Thomas},
  {Thoudam}, {Toribio}, {van der Tol}, {Usov}, {van Veelen}, {van der Veen},
  {ter Veen}, {Verbiest}, {Vermeulen}, {Vermaas}, {Vocks}, {Vogt}, {de Vos},
  {van der Wal}, {van Weeren}, {Weggemans}, {Weltevrede}, {White}, {Wijnholds},
  {Wilhelmsson}, {Wucknitz}, {Yatawatta}, {Zarka}, {Zensus}, \& {van
  Zwieten}}]{vanHaarlem13}
{van Haarlem}, M.~P., {Wise}, M.~W., {Gunst}, A.~W., {et~al.} 2013, \aap, 556,
  A2

\bibitem[{{van Weeren} {et~al.}(2021){van Weeren}, {Shimwell}, {Botteon},
  {Brunetti}, {Br{\"u}ggen}, {Boxelaar}, {Cassano}, {Di Gennaro},
  {Andrade-Santos}, {Bonnassieux}, {Bonafede}, {Cuciti}, {Dallacasa}, {de
  Gasperin}, {Gastaldello}, {Hardcastle}, {Hoeft}, {Kraft}, {Mandal},
  {Rossetti}, {R{\"o}ttgering}, {Tasse}, \& {Wilber}}]{vanWeeren21}
{van Weeren}, R.~J., {Shimwell}, T.~W., {Botteon}, A., {et~al.} 2021, \aap,
  651, A115

\bibitem[{{Vernstrom} {et~al.}(2021){Vernstrom}, {Heald}, {Vazza}, {Galvin},
  {West}, {Locatelli}, {Fornengo}, \& {Pinetti}}]{Vernstrom21}
{Vernstrom}, T., {Heald}, G., {Vazza}, F., {et~al.} 2021, \mnras, 505, 4178

\bibitem[{{Voges} {et~al.}(1999){Voges}, {Aschenbach}, {Boller},
  {Br{\"a}uninger}, {Briel}, {Burkert}, {Dennerl}, {Englhauser}, {Gruber},
  {Haberl}, {Hartner}, {Hasinger}, {K{\"u}rster}, {Pfeffermann}, {Pietsch},
  {Predehl}, {Rosso}, {Schmitt}, {Tr{\"u}mper}, \& {Zimmermann}}]{Voges99}
{Voges}, W., {Aschenbach}, B., {Boller}, T., {et~al.} 1999, \aap, 349, 389

\bibitem[{{Wen} \& {Han}(2024)}]{WH2024}
{Wen}, Z.~L. \& {Han}, J.~L. 2024, \apjs, 272, 39

\end{thebibliography}

\begin{appendix}

\section{Subtraction of compact sources and related uncertainty on the flux density}
\label{app}

In this Appendix, we report additional information about the source subtraction procedure and the residual contribution of non-subtracted sources to the newly detected emission.

\begin{figure}     
\includegraphics[width=0.9\linewidth]{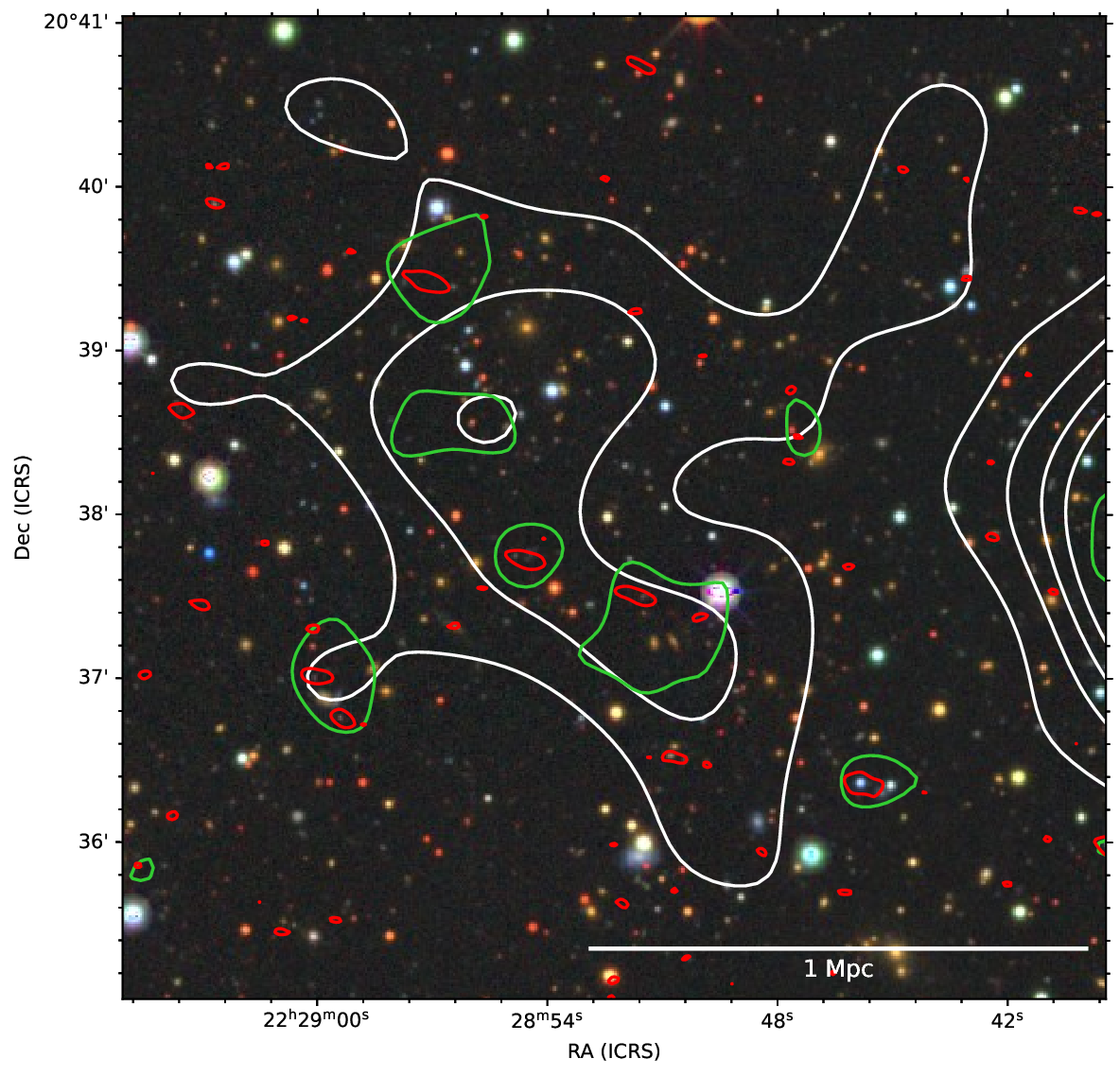}
\caption{Zoom-in on the eastern ‘‘ear’’ of the optical image obtained from the Legacy Surveys. Red contours show the 3$\sigma$ level of the LOFAR image created with an inner uv-cut of 2865$\lambda$, with restoring beam $4.5\arcsec\times10.6\arcsec$. White contours show the [3,5,7,9]$\times\sigma$ levels of the residual diffuse emission with restoring beam $69\arcsec\times81\arcsec$. Green contours show the 3$\sigma$ level of the total emission, with a restoring beam of $27\arcsec\times29\arcsec$. }
\label{fig:optical-hres-E}
\end{figure}

\begin{figure}     
\includegraphics[width=0.9\linewidth]{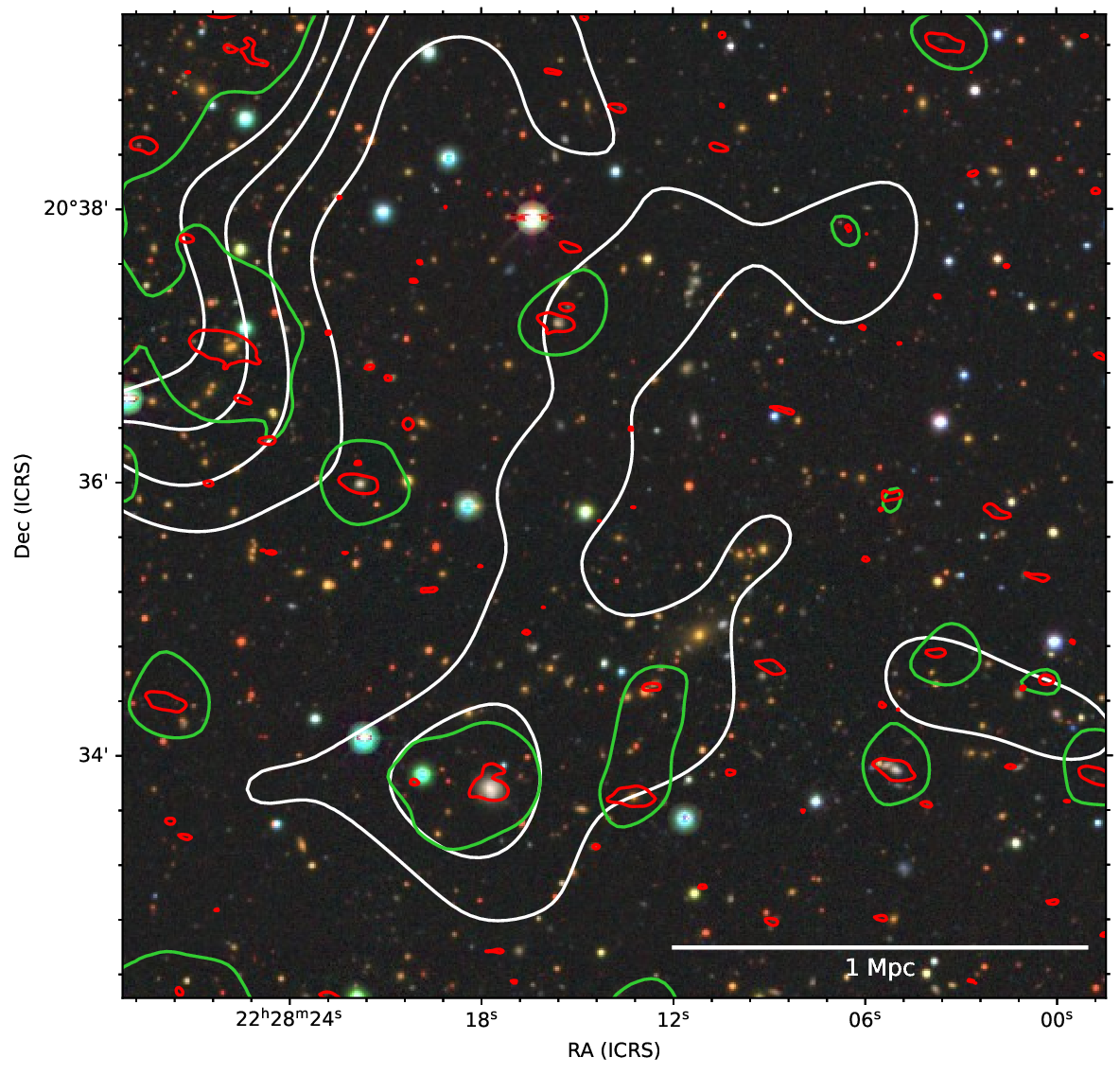}
\caption{Zoom-in on the south-western ‘‘ear’’ of the optical image obtained from the Legacy Surveys. Contours are as in Fig.\ref{fig:optical-hres-E}. See Tab.\ref{tab:radioimages} for details about the radio images.}
\label{fig:optical-hres-SW}
\end{figure}

The zoom-in of the optical image with radio contours overlaid is shown in Fig.~\ref{fig:optical-hres-E} and \ref{fig:optical-hres-SW}. Red contours show the image used for the subtraction of point-like sources, white contours show the residual diffuse emission and green contours show the total emission at an intermediate resolution. This overlay testifies to the presence of faint diffuse emission that is not strictly connected to compact sources. However, we can also see that some residuals from subtraction may contribute to shaping the faint extended emission. This contribution should also be considered in the total uncertainty of the measured flux density. 

The uncertainty on the flux density given in Sec.~\ref{sec:radio} comprises both the statistical error related to the rms noise and the integration area, the systematic error due to uncertainty of the calibration \citep[$10\%$ of the flux, see][]{Shimwell22}, and another systematic error due to the unperfected subtraction of compact sources.

We estimated the latter contribution following the methodology described by \citet{Botteon22}, who proposed an empirical approach based on the analysis of $\sim$300 galaxy clusters (see their Section 5). We computed the total flux density of discrete radio sources from the uv-cutted image ($\lambda$>2865) within the regions of the east and south-western ‘‘ears’’. In both areas, this is less than 10 mJy (4.7 and 6.2 mJy, respectively). An uncertainty equal to the 16$\%$ of this flux density was added in quadrature to the statistical error and the $10\%$ calibration error.

Furthermore, we would like to estimate the potential contribution to the flux density of the two ‘‘ears’’ from faint compact sources which could not have been subtracted because they are below the 3$\sigma$ level of the uv-cutted image. We conducted the following test. We modified the model derived from the cleaning of the uv-cutted image by dividing the model image by a factor of 300 and rotating it by $90^{\circ}$ and $180^{\circ}$. The scaling factor was selected such that the brightest source within the cluster field (peaking at 104 mJy/beam) was reduced to below the 3$\sigma$ level of the image (0.36 mJy/beam). This procedure generated two additional models containing a population of faint compact sources. These sources mimic the spatial distribution of real sources, showing an increasing concentration toward the cluster center, but all lie below the 3$\sigma$ thresholds of the image. 

We added these two models to the previously uv-subtracted visibilities and created two new images using the same parameters as those used for the $\lambda$>80 and $60\arcsec$ uv-tapered image (Fig.\ref{fig:X-optical}, bottom right). In these new images, the noise level remained unchanged (as all added sources were below the noise threshold), but the flux density of the two ‘‘ears’’ was intentionally increased.

For the E ‘‘ear’’ the flux density in the original image is 20.36 mJy. With the faint source model injected at 90$^{\circ}$, the flux density increased to 20.79 mJy, while with the model injected at $180^{\circ}$, it increased to 20.93 mJy. Similarly, for the SW ‘‘ear’’ the original flux density of 18.01 mJy rose to 18.19 mJy with the $90^{\circ}$-rotated model and 18.03 mJy with the $180^{\circ}$-rotated model. Thus, the maximum observed increase in flux density is approximately $3\%$ of the measured flux. Although this is a simplified test, the results suggest that the contribution of un-subtracted compact sources is likely of this magnitude. The statistical error, computed as the rms noise multiplied by the square root of the number of beams, is 2 mJy for both the ‘‘ears’’, thus the systematic effect is lower than the statistical error and the $10\%$ calibration error.

\end{appendix}


\end{document}